\documentclass[prl,twocolumn,showpacs,superscriptaddress]{revtex4}
\usepackage{graphicx}
\usepackage{dcolumn}
\usepackage{amsmath}
\usepackage{amssymb}
\usepackage{amsmath,verbatim}
\usepackage{bm}
\def\rv{{\bf r}}

\begin{document}
\title{Can time-dependent density functional theory predict 
the excitation energies of conjugated polymers?}
\author{Jianmin Tao}
%\email{jtao@lanl.gov}
\affiliation{Theoretical Division and CNLS,
Los Alamos National Laboratory, Los Alamos, New Mexico 87545}
\author{Sergei Tretiak}
\affiliation{Theoretical Division and CNLS,
Los Alamos National Laboratory, Los Alamos, New Mexico 87545}
\affiliation{Center for Integrated
Nanotechnology, Los Alamos National Laboratory, Los Alamos, New
Maxico 87545}
\author{Jian-Xin Zhu}
\affiliation{Theoretical Division and CNLS,
Los Alamos National Laboratory, Los Alamos, New Mexico 87545}

\date{\today}
\begin{abstract}
Excitation energies of light-emitting organic conjugated 
polymers have been investigated with time-dependent density functional 
theory (TDDFT) within the adiabatic approximation for the dynamical 
exchange-correlation potential. Our calculations show that the accuracy of 
the calculated TDDFT excitation energies largely depends upon the accuracy
of the dihedral angle obtained by the geometry optimization on 
ground-state DFT methods. We find that, when the DFT torsional dihedral angles 
between two adjacent phenyl rings are close to the experimental dihedral 
angles, the TDDFT excitation energies agree fairly well with experimental
values.  Further study shows that, while hybrid density functionals can 
correctly respect the thumb rule between singlet-singlet and singlet-triplet
excitation energies, semilocal functionals do not,
suggesting inadequacy of the semilocal functionals in 
predicting triplet excitation energies of conjugated polymers. 
\end{abstract}
\maketitle

%\section{Introduction}
%{\it Introduction:} 
The most important progress made in the development of molecular electronics is
the discovery of electroluminescent conjugated 
polymers~\cite{thomas98} -- that is, fluorescent 
polymers that emit light when these polymers in the excited states are stimulated
by, say, electric current. Conjugated polymers are organic semiconductors with
delocalized $\pi$-molecular orbitals along the polymeric chain. These materials are 
a major challenge to inorganic materials which have been dominating the commercial
market in light-emitting diodes for display and other purpose~\cite{pmay}. The 
attraction of conjugated polymers lies at their versatility, because their physical
properties such as color purity and emission efficiency can be fine-tuned by 
manipulation of their chemical structures. The systematic modification of the
properties of emissive polymers by synthetic design has become a vital component 
in the optimization of light-emitting devices. 

Theoretical investigation of their optical absorption plays a significant role
in computer-aided design and optimization of the electroluminescent polymers.
The method of choice for the simulation of the optical absorption of electronic
materials is time-dependent density functional theory (TDDFT)~\cite{grossbook}, 
owing to its high computational efficiency and comparable accuracy. TDDFT is 
the most important extension of Kohn-Sham ground-state DFT, the standard 
method in electronic structure calculations. The only approximation made 
in TDDFT is the dynamical exchange-correlation (XC) potential, which includes
all unknown many-body effects. The simplest construction 
is called adiabatic (ad) approximation~\cite{zs}, which takes the
same form of the static XC potential but replaces the ground-state density $n_0(\rv)$
with the instantaneous time-dependent density $n(\rv,t)$: 
%\begin{eqnarray}\label{adiabatic}
$v_{\rm xc}^{ad}([n];\rv,t) =
\delta E_{\rm xc}[n_0]/\delta n_0(\rv)
|_{n_0(\rv) = n(\rv,t)}~.$
%\end{eqnarray}
The advantage of this 
approach is its simplicity in both theoretical construction and numerical 
implementation. Although the adiabatic TDDFT cannot properly describe 
multiple excitations, it has become the most popular approach in the study 
of low-lying single-particle excitations (i.e., only one electron in the 
excitated states).

%%%%%%%%%%%%%%%%%%%%%%%%%%%%%%%%%%%%%%%%%%%%%%%%%%%%%%%%%%%%%%%%%
\begin{figure}
\includegraphics[width=\columnwidth]{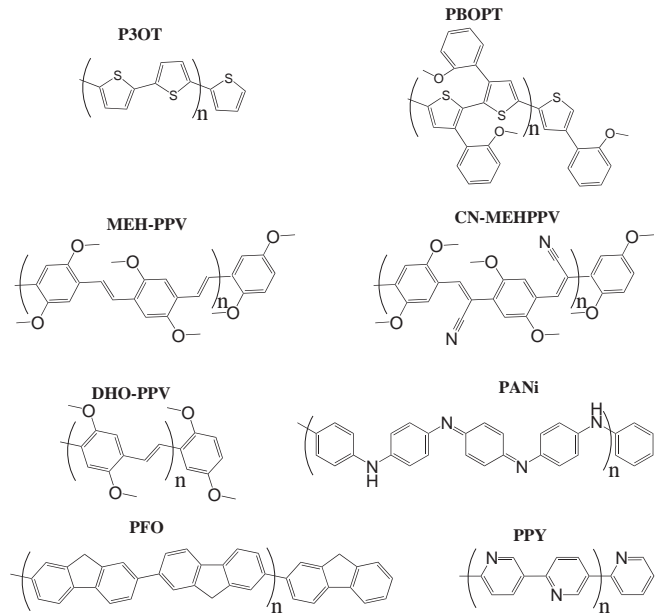}
\caption{Chemical structures of the computationally studied light-emitting
congugated polymers.}
\label{fpolymers}
\end{figure}
%%%%%%%%%%%%%%%%%%%%%%%%%%%%%%%%%%%%%%%%%%%%%%%%%%%%%%%%%%%%%%%%%

%%%%%%%%%%%%%%%%%%%%%%%%%%%%%%%
\begin{table*}
%\begin{sidewaystable}
\caption{Excitation energies of singlet-singlet ($S_0-S_1$) and singlet-triplet
($S_0-T_1$) gaps (in units of eV) of polymers of length of $ \sim 10~ {\rm nm}$
in gas phase calculated using the adiabatic TDDFT methods with the ground-state
geometries optimized on the respective density functionals. Basis set 6-31G is used in all
calculations.  The number in parentheses is the number of rings included in our
calculations. 1 hartree = 27.21 eV.}
%\begin{ruledtabular}
%\begin{tabular}{ldddddddddddd}
%& \multicolumn{5}{c}{$S_0-S_1$} & & \multicolumn{5}{c}{$S_0 - T_1$}
%\\ \cline{2-6} \cline{8-12}
%Polymer&
%\multicolumn{1}{c}{LSDA} & \multicolumn{1}{c}{TPSS} &
%\multicolumn{1}{c}{TPSSh} &
%\multicolumn{1}{c}{B3LYP} & \multicolumn{1}{c}{PBE0} &
%& \multicolumn{1}{c}{ LSDA} & \multicolumn{1}{c}{TPSS} &
%\multicolumn{1}{c}{TPSSh} &
%\multicolumn{1}{c}{B3LYP} & \multicolumn{1}{c}{PBE0} \\ \hline
\begin{ruledtabular}
\begin{tabular}{llddddddlddddddd}
& \multicolumn{6}{c}{$S_0-S_1$} & & \multicolumn{6}{c}{$S_0 - T_1^a$}
\\ \cline{2-7} \cline{9-14}
Polymer& \multicolumn{1}{c}{Expt$^a$}
 & \multicolumn{1}{c}{LSDA} & \multicolumn{1}{c}{TPSS} &
\multicolumn{1}{c}{TPSSh} &
\multicolumn{1}{c}{B3LYP} & \multicolumn{1}{c}{PBE0} &
& \multicolumn{1}{c}{ Expt$^a$}
& \multicolumn{1}{c}{ LSDA} & \multicolumn{1}{c}{TPSS} &
\multicolumn{1}{c}{TPSSh} &
\multicolumn{1}{c}{B3LYP} & \multicolumn{1}{c}{PBE0} \\ \hline

P3OT(28)     &2.8-3.8 &0.99 &0.99 &1.35&1.59 &1.76 &&1.7-2.2& 0.90 &0.80 &0.88&0.96 &0.95 \\
PBOPT(32)    &2.52 &1.49 &1.55 &1.96&2.26 &2.39 &&1.60&1.37 &1.31 &1.42&1.57 &1.54 \\
MEHPPV(16)   &2.48 &1.14 &1.27 &1.66&1.94 &2.07 &&1.30&1.04 &1.08 &1.18&1.31 &1.24 \\
PFO(36)      &3.22 &2.30 &2.45 &2.89&3.13 &3.30 &&2.30&2.22 &2.23 &2.34&2.45 &2.43 \\
DHOPPV(16)   &2.58 &1.14 &1.27 &1.67&1.95 &2.07 &&1.50&1.04 &1.08 &1.18&1.32 &1.24 \\
PPY(24)      &3.4-3.9 &1.82 &2.10 &2.61&2.87 &3.03 &&2.4-2.5&1.82 &1.99 &2.11&2.23 &2.20 \\
CN-MEHPPV(16)&2.72 &1.10 &1.34 &1.84&2.16 &2.27 &&N/A&1.06 &1.22 &1.34&1.48 &1.43 \\
PANi(20)     &2.00 &2.34 &2.53 &3.05&3.30 &3.44 &&$<0.9$&2.31 &2.43 &2.63&2.75 &2.73 \\
%\hline
\end{tabular}
\end{ruledtabular}

\footnotesize{$^a$From Ref.~\cite{monkman01}, in which 
there is a small red shift in gas phase, compared to those
in solvent (see discussion in the context).  $^b$Notation of Ref.~\cite{birks} is used.
Note that all the groups of -(CH$_2$)$_n$CH$_3$ in polymers have been replaced 
with the hydrogen (-H).}
%\end{sidewaystable}
\end{table*}
%%%%%%%%%%%%%%%%%%%%%%%%%%%%%%%%%%%%%%%%%%%%%%%%%%%%%%

Our previous studies of small molecules~\cite{ttz082} and 
molecular materials~\cite{tt09} show that 
the excitation energies obtained with the adiabatic TDDFT agree fairly well
with experiments. In the present work, we calculate the lowest singlet-singlet 
($S_0-S_1$) and singlet-triplet ($S_0 - T_1$) 
excitation energies of a series of light-emitting 
organic conjugated polymers (see Fig. 1 for their chemical structures).
The singlet-singlet excitation is responsible for the strong 
ultraviolet (UV) or 
near-UV optical absorption, while the singlet-triplet excitation
is responsible for weak fluorescence. Our calculations show that, 
when the dihedral angles~\cite{note1} between two adjacent phenyl rings obtained 
by the geometry optimization on ground-state DFT methods
are close to experimental dihedral angles, the calculated TDDFT 
excitation energies agree well with experiments, regardless of whether 
the excitations arise from singlet-singlet excitations or 
singlet-triplet excitations. This suggests that 
in TDDFT calculations, there are two
sources of error. One is from the adiabatic approximation itself~\cite{note2},  
and the other, much larger than the first one, arises from inaccuracy of 
the ground-state DFT geometries. In order to identify these errors, here
we employ five commonly-used density functionals. Two of them, 
the local spin density approximation (LSDA) and the meta-generalized 
gradient approximation (meta-GGA) of Tao, Perdew, Staroverov, and 
Scuseria (TPSS)~\cite{tpss}, are pure density functionals, while
the other three, TPSSh~\cite{sstp1} (a hybrid of the TPSS meta-GGA
with $10\%$ exact exchange), B3LYP~\cite{b3lyp} (a hybrid with $20\%$ exact exchange), and
PBE0~\cite{gus99} (a hybrid of the Perdew-Burke-Ernzerhof (PBE)~\cite{pbe96} GGA 
with $25\%$ exact exchange) are hybrid functionals with increasing amount 
of exact exchange from TPSSh, B3LYP to PBE0.

Moreover, in the simulation of electronic excitations of small 
molecules and molecular materials, the most effort has been devoted 
to the study of the absorption arising from singlet-singlet excitation, 
leaving the singlet-triplet excitation less investigated~\cite{perun}. 
An important reason for this omission is that triplet-state 
energies are not easy to measure through direct optical absorption due 
to very low singlet-triplet ($S_0 - T_1$) absorption 
coefficient~\cite{walters} and low phosphorescence quantum yield~\cite{roman} 
($<10^{-6}$). The major approaches to 
probe triplet states in conjugated polymers are the charge recombination 
or energy transfer, and singlet-triplet ($T_1 - S_0$ or $S_1 - T_1$) 
intersystem crossing~\cite{reindl,parker,birks}.
The observation of $T_1 - S_0$
phosphorescence from molecules initially excited into $S_1$ is clear evidence
for a radiationless transition from $S_1$ to an isoenergetic level of the
triplet manifold, corresponding to {\it singlet-triplet intersystem crossing}.
Singlet-triplet intersystem crossing can occur either from the zero-point
vibrational level of $S_1$ or from thermally-populated vibrational level of 
$S_1$ into an excited vibrational level of $T_1$, or more probably into
a higher excited triplet state $T_2$, which is closer in energy to $S_1$.
It has been found~\cite{burrows01,burrows02} that the properties of the
triplet states directly impact device performance. For example,
the formation of triplet states may cause the loss of the device efficiency in these
materials and thus can limit device performance and operational life span. Therefore,
investigation of triplet excitations is crucial for a full 
understanding of electroluminescence behavior of conjugated organic polymers
and for the improvement of new materials.

Monkman and collaborators~\cite{burrows02,monkman01} investigated the photophysics
of triplet states in a series of conjugated polymers and measured
the excitation energies of the lowest singlet- and triplet-excitated states. Their
measurements show that the excitation energies in general respect the well-known
rule of thumb found for small molecules: 
\begin{eqnarray}\label{thumb}
E_{T} \approx 2E_{S}/3, 
\end{eqnarray}
where $E_T$ is the triplet excitation energy
and $E_S$ is the singlet-singlet excitation energy. As a second part of
our work, we calculate the singlet-triplet excitation energies of the 
polymers with the adiabatic TDDFT. We find that, without exact exchange mixing, a pure
semilocal density functional cannot satisfy the thumb rule of Eq.~(\ref{thumb}),
suggesting inadequacy of the adiabatic semilocal functionals in predicting the
triplet excitation energies for polymers.

%%%%%%%%%%%%%%%%%%%%%%%%%%%%%%%%%%%%%%%%%%%%%%%%%%%%%%
\begin{table}
\caption{Torsions of the conjugated polymers}
\begin{tabular}{llccccc|l}
\hline \hline
 Polymer&Expt & PBE0 & Energy \\ \hline
P3OT      &$\sim 24\,^{\circ}$ & $\sim 0\,^{\circ}$&red shift \\
PBOPT     &$\sim 35\,^{\circ}$ &$\sim 40\,^{\circ}$ &On experiment \\
MEHPPV    &$\sim 20\,^{\circ}$ &$\sim 1\,^{\circ}$ &red shift\\
PFO       &$\sim 40\,^{\circ}$ &$\sim 38\,^{\circ}$ &On experiment \\
DHOPPV    &$\sim 20\,^{\circ}$ &$\sim 0\,^{\circ}$ &red shift\\
PPY       & $\gtrsim 0\,^{\circ}$&$\sim 0-1\,^{\circ}$&slightly red shift\\
CN-MEHPPV &$\sim 20\,^{\circ}$ &$\sim 0\,^{\circ}$ &red shift\\
PANi      &$\sim 0\,^{\circ}$ & $\sim 18-26\,^{\circ}$&too blue shift\\
\hline \hline
\end{tabular}

%\footnotesize{$^a$From Ref.~\cite{nielsen80}.}
\end{table}
%%%%%%%%%%%%%%%%%%%%%%%%%%%%%%%%%%%%%%%%%%%%%%%%%%%%%%%%%%%
\begin{table*}
\caption{Excitation energies of singlet-singlet ($S_0-S_1$) and singlet-triplet
($S_0-T_1$) gaps (in units of eV) of polymers of length of
$ \sim 10~ {\rm nm}$ in benzene solution calculated using the adiabatic
TDDFT methods with the ground-state geometries optimized on the respective
density functionals. The solvent effects are taken into account through
PCM (polarizable continuum model) method. Basis set 6-31G is used in all
calculations. The number in parentheses is the number of rings included in our
calculations.
1 hartree = 27.21 eV.}
\begin{ruledtabular}
\begin{tabular}{llddddddlddddddd}
& \multicolumn{6}{c}{$S_0-S_1$} & & \multicolumn{6}{c}{$S_0 - T_1^{b}$}
\\ \cline{2-7} \cline{9-14}
Polymer& \multicolumn{1}{c}{Expt$^a$}
 & \multicolumn{1}{c}{LSD} & \multicolumn{1}{c}{TPSS} &
\multicolumn{1}{c}{TPSSh} &
\multicolumn{1}{c}{B3LYP} & \multicolumn{1}{c}{PBE0} &
& \multicolumn{1}{c}{ Expt$^a$}
& \multicolumn{1}{c}{ LSD} & \multicolumn{1}{c}{TPSS} &
\multicolumn{1}{c}{TPSSh} &
\multicolumn{1}{c}{B3LYP} & \multicolumn{1}{c}{PBE0} \\ \hline

P3OT(28)     &2.8-3.8 &0.97 &0.97 &1.32 &1.56 &1.73 &&1.7-2.2 &0.89 &0.80 &0.87 &0.95 &0.94 \\
PBOPT(32)    &2.52 & & & & & &&1.60 & & & & & \\
MEHPPV(16)   &2.48 &1.12 &1.25 &1.64 &1.91 &2.04 &&1.30 &1.03 &1.07 &1.18 &1.32 &1.25 \\
PFO(36)      &3.22 &2.30 &2.45 &2.88 &3.12 &3.29 &&2.30 &2.22 &2.24 &2.35 &2.46 &2.43 \\
DHOPPV(16)   &2.58 &1.12 &1.25 &1.64 &1.92 &2.04 &&1.50 &1.03 &1.07 &1.18 &1.32 &1.25 \\
PPY(24)      &3.4-3.9 &2.08 &2.16 &2.61 &2.85 &3.01 &&2.4-2.5 &2.02 &1.99 &2.11 &2.23 &2.20 \\
CN-MEHPPV(16)&2.72 &1.10 &1.32 &1.80 &2.10 &2.21 &&N/A &1.05 &1.21 &1.34 &1.48 &1.43 \\
PANi(20)     &2.00  &2.33 &2.53 &3.03 &3.27 &3.41 &&$<0.9$&2.30 &2.42 &2.62 &2.75 &2.73 \\

\hline
\end{tabular}
\end{ruledtabular}

\footnotesize{$^a$From Ref.~\cite{monkman01}. 
$^b$Notation of Ref.~\cite{birks} is used.  
Note that all the groups of -(CH$_2$)$_n$CH$_3$ in polymers have been replaced 
with the hydrogen (-H).} 
\end{table*}
%%%%%%%%%%%%%%%%%%%%%%%%%%%%%%%%%%%%%%%%%%%%%%%%%%%%%%%%%%%%%%%%%%%%%%%%%%

{\it Computational method:}
All our calculations were performed on the molecular-structure code Gaussian 03~\cite{g03}. The
initial geometries are prepared with GaussView 4, while the dihedral angles are manually
adjusted to be $\sim 30\,^{\circ}$. Then we optimize the geometries on respective 
ground-state DFT methods. Finally we calculate the excitation energies from the 
optimized ground-state geometries with the adiabatic TDDFT density functionals. For
consistency, basis set 6-31G was used in both ground-state and time-dependent DFT 
calculations. In order to check whether our conclusion is affected by the choice 
of basis set, we repeat our calculations for polymer P3OT using a 
larger basis set 6-31G(d) that has diffusion functions. Our calculations show that
the excitation energy obtained with 6-31G(d) is larger only by $<0.2$ eV than that 
obtained with 6-31G basis set.   
The excitation energies of the polymers in 
benzene solvent are calculated with PCM (polarizable continuum model)~\cite{cmt97}.  
The polymers we study here have chain length of $\sim 10$ nm. 
Since the groups of -(CH$_2$)$_n$CH$_3$ only has little effect upon the properties
of the polymers~\cite{ttz082}, these groups have been removed from the 
backbone of a polymer and are, therefore, excluded in all calculations.

%{\it Results and discussion:} 
Table 1 shows the first singlet and triplet excitation energies 
of the polymers in gas phase calculated with the 
adiabatic TDDFT. The experimental results
are also listed for comparison. Usually a polymer is of infinite chain length.
In practical calculations, we only choose several repeating monomeric units.
The number of ``molecular'' rings included in our calculations for each polymer
is given in the parentheses in Tables 1 and 3. 
These numbers are chosen so that the lengths of the
polymers are about 10 nm. This size effect will be reduced by increasing
the repeating units. However, adding the  
repeating units will simultaneously increase the computational time.
On the other hand, high accuracy usually can be achieved by using large
basis set, which will result in significant increase in computational
time. In the present calculations, we use a basis set which is relatively smaller 
than those used in small molecular calculations, and prepare the polymers 
with moderate length of chain. This is a balanced choice between  
the size effect and the accuracy we can tolerate. 

From Table 1
we observe that, among the five adiabatic TDDFT methods, the adiabatic 
PBE0 functional yields the most
accurate excitation energies. This is consistent with our previous 
studies~\cite{ttz082,tt09}. We can see from Table 1 that the difference between the
singlet and the triplet excitation energies, $E_S-E_T$, is $\sim 0-0.1$ eV for LSDA, 
$\sim 0.1-0.2$ eV for meta-GGA, $\sim 0.5$ eV for TPSSh, $\sim 0.6$ eV for B3LYP, 
and $\sim 0.8$ eV for PBE0. The difference increases as the amount of exact exchange
increases. However, some studies suggest~\cite{itc05,itc07} that 
for semilocal density functionals (LSDA, GGA, and meta-GGA), 
this difference may vanish in the limit of infinite chain length,
a result similar to the performance of semilocal functionals for solids. Mixing
exact exchange into a semilocal functional will partly correct the errors
from self interaction, improve the asymptotic behavior of the XC potential, and   
build in other many-body properties such as excitonic effects~\cite{itc05,itc07} 
which have not been taken into account properly in pure density functional 
approximations and thus will lead to a finite difference in this limit.

Interestingly, we find that when the theoretical dihedral angle is
smaller than the experimental dihedral angle, the TDDFT methods tend
to underestimate the excitation energies regardless of whether the
excitation is singlet or triplet. When the theoretical dihedral 
angle is close to the experimental one, the TDDFT excitation energies
are in good agreement with experiments. Our calculations show that in rare cases, 
theoretical dihedral angles can be greater that experimental estimates.
In this case, the excitation energies are overestimated by the TDDFT
methods. A comparison of the dihedral angles between theoretical and 
experimental estimates is displayed in Table 2. The origin of torsional
angles (or generally tortional disorder) of polymers is complicated. 
It may arise from interchain interaction in amorphous polymeric
materials~\cite{sergei1,sergei2} or from van der Waals 
interaction~\cite{dion,scheffler}
between phenyl rings. These effects have not been properly taken into 
account in current DFT methods.

The excitation energies of the polymers in benzene solvent are summarized in Table 3. 
From Table 3, we can see that the lowest singlet-singlet excitation energies 
in solution have a red shift of $\sim 0.01-0.05$ eV, compared to those in gas 
phase (Table 1). This is consistent with what we have observed for 
oligomers~\cite{ttz082,tt09}. However, this trend does not apply to the triplet 
excitation. Triplet excitation energies are nearly the same whether the 
polymer is in gas phase or in solution. 

%{\it Conclusion:}
In conclusion, we have investigated the lowest excitation energies of several
light-emitting conjugated polymers from the adiabatic TDDFT methods. Our 
calculations show that the calculated excitation energies are in good aggrement
with experiments only when the theoretical torsions agree with experimental 
estimates. If the theoretical dihedral angles are smaller than the experiments,
the TDDFT excitation energies tend to be underestimated. If the theoretical dihedral 
angles are greater than the experiments, as in rare case, the TDDFT excitation 
energies tend to be overestimated. Furthermore, we find that, a 
semilocal functional without exact exchange mixing does not satisfy
the well-known ``two-third'' thumb rule relation between the singlet-singlet and 
singlet-triplet excitation energies. For semilocal functionals,
the difference in energy between singlet state
and triplet state is less than 0.1 eV for polymers with chain length of 10 nm and 
may vanish in the limit of infinite chain length. 
Compared to semilocal functionals, hybrid functionals yield
much larger difference between singlet-singlet and singlet-triplet excitation 
energies for polymers with finite chain length as well as with infinite chain length.
This difference increases with more exact exchange mixed in
semilocal functionals, and is nonzero even in the limit of infinite chain length.

\acknowledgments
The authors thank Richard Martin and John Perdew for valuable discussion and suggestions. 
This work was carried out under the auspices of the National Nuclear Security Administration
of the U.S. Department of Energy at Los Alamos National Laboratory under Contract No. 
DE-AC52-06NA25396, and was supported by the LANL LDRD program.

%%%%%%%%%%%%%%%%%%%%%%%%%%%%%%%%%%%%%%%%%%%%%%

\begin{thebibliography}{100}
\bibitem{note1}
Dihedral angle is defined as the angle formed by two planes containing two
adjacent phenyl or other ``molecular'' rings.
\end{thebibliography}
\end{document}